\documentstyle[pra,aps]{revtex}
\tightenlines
\RequirePackage{epsfig}
\begin{document}
\draft
\title{Toroidal Optical Dipole Traps for Atomic Bose-Einstein
       Condensates using Laguerre-Gaussian Beams}
\author{E. M. Wright,
        \footnote{\vspace {0.5cm}
        Permanent address: Optical Sciences Center, University
        of Arizona, Tucson, AZ 85721, USA}
        J. Arlt, and K. Dholakia\\
        School of Physics \& Astronomy\\
      University of St. Andrews, North Haugh\\
        St. Andrews, Fife KY16 9SS
        Scotland, UK}

\date{\today}
\tolerance = 1000

\newcommand{\MHz}{\:\text{MHz}}
\newcommand{\kHz}{\:\text{kHz}}
\newcommand{\nm}{\:\text{nm}}
\newcommand{\um}{\:\mu\text{m}}
\newcommand{\mm}{\:\text{mm}}
\newcommand{\cm}{\:\text{cm}}
\newcommand{\mW}{\:\text{mW}}
\newcommand{\ISat}{I_{\text{Sat}}}
\newcommand{\nK}{\:\text{nK}}
\maketitle
\begin{abstract}
We theoretically investigate the use of red-detuned
Laguerre-Gaussian (LG) laser beams of varying azimuthal mode index
for producing toroidal optical dipole traps in two-dimensional
atomic Bose-Einstein condensates. Higher-order LG beams provide
deeper potential wells and tighter confinement for a fixed toroid
radius and laser power. Numerical simulations of the loading of
the toroidal trap from a variety of initial conditions is also
given.
\end{abstract}
\newpage
\section{Introduction}
Recent work has seen unprecedented advances in the preparation of
Bose-Einstein condensates (BEC) of dilute alkali vapors
\cite{AndEnsMat95,DavMewAnd95,DalGioPit99}. These
quantum degenerate systems have paved the way for numerous
innovative studies of weakly interacting Bose gases. An important
area within this field is the generation and study of quantized
vortices on atomic mesoscopic rings and the potential to study
effects due to persistent
currents and Josephson effects
\cite{Rok97,JavPaiYoo98,BusAng99,PetYou99,BenRagSme99,GelGol00,IsoNakOhm00}.
Central to such studies is investigation of geometries for
generation of a BEC in a toroidal trap \cite{SalParRea99,CloseZ99}.
Typically BEC is created in
a magnetic trap that confines atoms in weak-field seeking states
and is thus dependent on the atomic hyperfine states. A toroidal
trap can then be formed by piercing a magnetic trap with a
blue-detuned laser at its center \cite{DavMewAnd95}.
However, the state dependence of the trapping is a
limitation for advanced studies including multi-component spinor
condensates. Further, the magnetic field is primarily dictated by
the trapping requirements setting limitations on its spatial form
and amplitude. Optical dipole traps can potentially circumvent
these problems being state independent \cite{StaAndChi98}.
Importantly, the spatial
form of the optical trap is dictated by the light beams used. This
allows one to potentially generate arbitrary shapes of condensate
\cite{Dur99}.
Experimental work has shown the ability to transfer a condensate
from a magnetic trap to an optical dipole trap created from a
tightly focused Gaussian light beam \cite{StaAndChi98}.

The circularly symmetric Laguerre-Gaussian (LG) laser modes have
generated substantial interest in recent years. This stems from
the identification that they possess an orbital angular momentum
of $\ell\hbar$ per photon \cite{Les92}. This is in addition to the
spin angular momentum associated with the polarization state of
the beam. The azimuthal index $\ell$ refers to the number of $2\pi$
phase cycles around the mode circumference. A given mode has
$p+1$ radial nodes in the mode profile where $p$ is the other
index used in the LG mode description. One finds that LG beams
with radial index $p=0$ are in the form of an annulus
\cite{CliffordACD98}. There are several techniques for production
of LG modes including the use of a cylindrical lens mode converter
\cite{Beij93} and holographic methods \cite{He95}. The latter
technique allows generation of LG modes from the fundamental
output of a laser beam.  The annulus becomes thinner as one
increases the azimuthal index $\ell$ of the mode \cite{CliffordACD98}.

In this work we present a technique for generating a toroidal BEC
employing a Laguerre-Gaussian light beam. Single-ringed forms of
such beams offer for the first time a direct possibility of
generating ring-shaped condensates in an optical potential. We
study the operation of this dipole trap as a function of azimuthal
mode index and discuss the loading of condensates into
such optical traps from a variety of initial conditions.
\section{Basic Model and Equations}
\subsection{Gross-Pitaevskii equation}
The basic model we consider is shown in Fig. 1
and comprises a two-dimensional
(2D) BEC whose normalized mode profile is frozen by tight confinement along
the $z$-direction, and which is placed at the focus of an off-resonant
LG laser beam of frequency $\omega_L$ incident along the $z$-axis that
provides a two-dimensional optical potential \cite{StaAndChi98}.
The tight confinement may be provided, for example, by an independent sheet
potential using a scanned optical dipole potential \cite{Dur99}.
Then the Gross-Pitaevskii equation (GPE)
describing the macroscopic wavefunction $\psi({\bf r}_\perp,t)$
for the quasi-2D motion can be written as \cite{KimWonOh00,PetHolShl00}
\begin{equation}
i\hbar\frac{\partial\psi}{\partial t} =
-\frac{\hbar^2}{2M}\nabla_\perp^2\psi
+ U({\bf r}_\perp)\psi + gN|\psi|^2\psi  \, ,
\label{GPeq1}
\end{equation}
where ${\bf r}_\perp$ is the two-dimensional position vector in the plane
perpendicular to $z$, $\nabla_\perp^2$ the corresponding two-dimensional
Laplacian,
$M$ is the atomic mass, and $g$ is the effective short-range interaction
strength. The potential term on the right-hand-side
\begin{equation}
U({\bf r}_\perp) =
\frac{\hbar\Gamma^2}{8\Delta}\left (\frac{I({\bf r}_\perp)}
{\ISat}\right )
\label{OptPot}
\end{equation}
describes the 2D optical dipole potential with
$\Delta=\omega_L-\omega_A$ the laser detuning
from the optical transition frequency $\omega_A$,
$\Gamma$ the natural linewidth of the
optical transition, $\ISat$ is the resonant saturation intensity, and
$I({\bf r}_\perp)=\frac{1}{2}\epsilon_0 c n |E({\bf r}_\perp,0)|^2$,
assuming the BEC is centered at $z=0$ and has a thickness less than the
Rayleigh range of the focused beam.
In addition, we assume that the intensity profile of
the LG beam is undistorted
upon propagation through the 2D BEC, which is reasonable for a thin BEC.
We remark, however, that the LG beam becomes spatially phase-modulated
upon transiting the BEC, resulting in a changed far field profile of the
beam that could prove a useful diagnostic of the BEC density profile.

For later purposes we point out the GPE conserves both the
wave function norm
\begin{equation}
n(t) = \int_{ }^{ } d^2 {\bf r}_\perp |\psi({\bf r}_\perp,t)|^2  ,
\label{Norm}
\end{equation}
and also the effective single-particle Hamiltonian
\begin{equation}
H(t) = \int_{ }^{ } d^2 {\bf r}_\perp \left [
\frac{\hbar^2}{2M} |\nabla\psi|^2 + U({\bf r}_\perp)
|\psi|^2 + \frac{gN}{2}|\psi|^4 \right ]  .
\label{Ham}
\end{equation}
\subsection{Toroidal optical dipole potential}
Our goal in this work is to explore theoretically the use of LG
beams for producing toroidal optical dipole traps, the dipole potential
being proportional to the laser intensity in the limit of large detunings
considered here.  The intensity profile of
an LG beam at its focus ($z=0$) in cylindrical coordinates
(${\bf r}_\perp=(r,\theta)$) takes the form
\begin{equation}
I_{p,\ell}(r) = \frac{2p!}{(p+|\ell|)!}
\frac{P_0}{\pi w_{p,\ell}^2}
\left (\frac{2r^2}{w_{p,\ell}^2} \right )^{|\ell|}e^{-2r^2/w_{p,\ell}^2}
\left [ L_p^{|\ell|}\left (\frac{2r^2}{w_{p,\ell}^2} \right )\right ]^2  ,
\end{equation}
where $P_0$ is the beam power, $\ell$ is the azimuthal
mode index, the field
having a variation $\exp(i\ell\theta)$, $p$ is the radial mode index which
is the number of radial intensity maxima, $w_{p,\ell}$ is the mode
spot size, and $L_p^{|\ell|}$ is a generalized Laguerre polynomial.
Here we create a single-ringed toroidal trap using a red-detuned laser
field ($\Delta<0$), so the atoms are light-seeking,
using an LG beam with $p=0$,
in which case $L_0^{|\ell|}=1$, and we hereafter drop the $p$ index
and take $\ell > 0$.
Then for each
value of $\ell$ the intensity profile has a single maximum at
$r_\ell = w_\ell\sqrt { \frac{\ell}{2} }$ \cite{ArlHitDho00}.
Thus, in order to produce a toroid of a fixed radius $r_\ell=r_T$, we need
to choose the spot size $w_\ell$ for each azimuthal mode such that
\begin{equation}
w_\ell = r_T\sqrt{ \frac{2}{\ell} } .
\label{well}
\end{equation}
Furthermore, for a given $\ell$ the peak intensity at the toroid center
$r=r_T$ may be written as
\begin{equation}
I_{\ell} = \frac{2P_0}{\pi w_\ell^2}
\left ( \frac{\ell^\ell e^{-\ell}}{\ell!} \right )
= I_1\left ( \frac{\ell^{\ell+1}e^{-(\ell-1)}}{\ell!}  \right )
\approx I_1 \sqrt{\ell}  ,
\end{equation}
with $I_1=e^{-1}P_0/\pi r_T^2$ the peak intensity for $\ell=1$
\cite{ArlHitDho00}. Here in the expression for $I_\ell$ we have used
Stirling's formula $\ell!\approx \sqrt{2\pi\ell}\cdot\ell^\ell/e^\ell$,
and we see that the peak intensity scales as $\sqrt{\ell}$.

Gathering the above results together,
for a fixed laser power $P_0$ we may write the optical dipole
potential (eq. (\ref{OptPot}))
in the following form which is
useful for comparison between different values of $\ell$
\begin{equation}
U_\ell(r) = U_\ell
\left (\frac{r}{r_T} \right )^{2\ell}
e^{-\ell (r^2/r_T^2-1)}  ,
\label{Uell}
\end{equation}
%
where
\begin{equation}
U_1 = \frac{\hbar\Gamma^2}{8\Delta}\left (\frac{e^{-1}P_0}
{\pi r_T^2\ISat}\right )  ,
\qquad
U_\ell = U_1\left (\frac{\ell^{\ell+1}e^{-(\ell-1)}}{\ell!} \right )
\approx U_1 \sqrt{\ell}  ,
\end{equation}
$U_1$ being the optical dipole potential well depth for
$\ell=1$ at the toroid radius $r_T$.
Note that $U_\ell$ is negative for a laser red detuned from resonance.
Figure 2 shows the normalized optical dipole potential $U_\ell(r)/|U_1|$ versus $r/r_T$
for various azimuthal mode indices $\ell$, a red-detuned laser,
and fixed laser power and toroid radius, i.e. fixed $U_1$.
Here we see that as $\ell$ increases the toroidal trap becomes
deeper and tighter, meaning that higher-order LG beams
present advantages for making tight toroidal dipole traps.
\section{Ground state toroidal solutions}
In this section we elucidate the ground state properties of toroidal
traps using the harmonic oscillator and Thomas-Fermi approximations.
These approximate solutions amply illustrate the key features of the problem.
\subsection{Thomas-Fermi solution}
Setting $\psi(r,\theta,t)=\phi(r)\exp(-i\mu t/\hbar)$ for the cylindrically
symmetric ground state we obtain

\begin{equation}
\mu\phi = -\frac{\hbar^2}{2M}\nabla_\perp^2\phi
+ U_\ell(r)\phi + gN|\phi|^2\phi  ,
\label{GPeq2}
\end{equation}
with $\mu$ the chemical potential.
To obtain insight into the solutions of this equation we consider the
harmonic oscillator approximation to the toroidal dipole potential
(eq. (\ref{Uell})) trap around $r=r_T$,
\begin{equation}
U_\ell(r_T+\delta r) \approx U_\ell\left
(1-2\ell\frac{\delta r^2}{r_T^2} + \ldots\right ) \, ,
\label{Hosc}
\end{equation}
where $U_\ell$ is negative for a red detuned laser, as assumed
throughout the paper.
%
If we furthermore employ the Thomas-Fermi approximation
\cite{EdwBur95,BayPet96}, in which
the kinetic energy term is neglected in comparison to the
mean field energy, then we obtain the approximate GPE
\begin{equation}
(\mu-U_\ell)\phi \approx \frac{1}{2}M\Omega_\ell^2\delta r^2\phi
+ gN|\phi|^2\phi
\label{GPeq3}
\end{equation}
where the effective harmonic oscillator frequencies are
\begin{equation}
\Omega_\ell = 2 \sqrt{ \frac{|U_1|}{Mr_T^2} }
\left (\frac{\ell^{\ell+2}e^{-(\ell-1)}}{\ell !} \right )^{1/2}
\approx 2 \sqrt{ \frac{|U_1|}{Mr_T^2} } \cdot\ell^{3/4}  ,
\label{Omell}
\end{equation}
where Stirling's formula was used in the last expression.
Equation (\ref{GPeq3}) has the approximate ring solution
$(r_T-\Delta r_\ell\le r \le r_T+ \Delta r_\ell)$
\begin{equation}
\mu = U_\ell + \frac{gN}{2\pi r_T \Delta r_\ell}  ,
\qquad
|\phi(r)|^2 =\frac{1}{2\pi r_T \Delta r_\ell}
\left ( 1 -\frac{(r-r_T)^2}{\Delta r_\ell^2} \right )  ,
\end{equation}
where the ring width is given by
\begin{equation}
\left ( \frac{\Delta r_\ell}{r_T} \right )=
\left ( \frac{gN}{4\pi |U_1|r_T^2} \right )^{1/3}
\left ( \frac{\ell !}{\ell^{\ell+2}e^{-(\ell-1)}} \right )^{1/3}
\approx \left ( \frac{gN}{4\pi |U_1|r_T^2} \right )^{1/3}
\ell^{-1/2}  .
\label{Delrell}
\end{equation}
These solutions correspond to ring-shaped BEC density
profiles which are
peaked at $r=r_T$ and have a width $2\Delta r_\ell$.
We see that as the azimuthal mode index $\ell$ is increased with
all other parameters fixed the radial harmonic oscillator frequency
increases $\Omega_\ell \propto \ell^{3/4}$, and the ring width
decreases $\Delta r_\ell\propto \ell^{-1/2}$.  Thus, using higher-order
LG beams provides an advantage for making tight bound toroidal
rings as commented earlier.

The 2D Thomas-Fermi solution above should be applicable under the
combined conditions $gN >> \hbar^2/2M$ and $\Delta r_\ell/r_T <<1$,
which ensure that the mean-field energy is greater than the kinetic
energy, and the ring width is less than the toroid radius.  The latter
condition is also required for the validity of the harmonic oscillator
approximation.
\subsection{Scaling}
To explore the parameter space for the toroidal trap it is useful to
introduce appropriately scaled units.  For a trap of radius $r_T$ we
scale all lengths as ${\bf \rho}_\perp={\bf r}_\perp/r_T$,
all energies are scaled
to $\hbar\omega_T=\hbar^2/Mr_T^2$, and we introduce a dimensionless time
$\tau=\omega_T t$.  In these units the GPE becomes
\begin{equation}
i\frac{\partial\varphi}{\partial\tau} = -\frac{1}{2}\nabla_\perp^2\varphi
+ u_\ell(\rho)\varphi + \pi\eta N|\varphi|^2\varphi  ,
\label{GPeq4}
\end{equation}
where $\varphi=r_T\psi$, $u_\ell=U_\ell/\hbar\omega_T$, and
$\eta=Mg/\pi\hbar^2$
is a dimensionless measure of the repulsive many-body interactions
\cite{KimWonOh00,PetHolShl00,BayTan98}.  We shall use the above scaled
GPE in our numerical simulations of loading toroidal optical dipole traps.

Turning to the Thomas-Fermi solution, in dimensionless units the radial
harmonic oscillator frequency and the ring width become
\begin{equation}
\frac{\Omega_\ell}{\omega_T}\approx 2\sqrt{|u_1|}\cdot\ell^{3/4}
\qquad
\frac{\Delta r_\ell}{r_T} \approx
\left (\frac{\eta N}{4|u_1|}\right )^{1/3}\ell^{-1/2}
\end{equation}
and the chemical potential is given by
\begin{equation}
\frac{\mu}{\hbar\omega_T}\approx u_1\ell^{1/2} \left [
1 + 2\cdot \text{sgn}(u_1)
\left (\frac{\eta N}{4|u_1|}\right )^{2/3}
\right ]
\label{ChemPot}
\end{equation}
where Stirling's formula has been used.
These solutions depend on the dimensionless well
depth $u_1$ for $\ell=1$ which is given explicitly by
\begin{equation}
u_1 = \frac{U_1}{\hbar\omega_T} =
\frac{\Gamma^2}{8\hbar\Delta}\left (\frac{e^{-1}MP_0}
{\pi \ISat}\right )  ,
\end{equation}
and is independent of the toroid radius.

In dimensionless form the condition $gN>>\hbar^2/2M$ for applicability
of the Thomas-Fermi approximation becomes $\eta N >> 1/(2\pi)$.
Combining this with the condition $\Delta r_\ell/r_T <1$ yields the
constraint on the number of particles
\begin{equation}
\frac{1}{2\pi} < \eta N < 4|u_1|\sqrt{\ell} ,
\end{equation}
which shows that traps using higher azimuthal mode index can
hold more atoms, all other parameters being equal.
\subsection{Parameter values}
To illustrate the basic scales involved in toroidal traps we
consider the case of the $\lambda_A=589$ nm transition of
Na using a laser wavelength of $\lambda_L=985 \nm$
and a toroid radius of $r_T=10 \um$, these values being
used throughout this paper.
Then using the parameter values
$\ISat=63$ W/m$^2$, $\Gamma=2\pi\times 9.89 \MHz$,
we obtain $u_\ell=u_1\sqrt{\ell}=-253\times P_0 \sqrt{\ell}$,
with $P_0$ in milli-Watts.
Using $\hbar\omega_T\equiv 2.1 \times 10^{-10}$ K, with
corresponding time scale $1/\omega_T=36$ ms, for a given azimuthal
mode index $\ell$ this potential
depth translates to 
$u_\ell = -53 \times P_0\sqrt{\ell} \nK$.
The corresponding effective oscillator frequencies in the
toroidal trap are then
$\Omega_\ell=2\pi\times 0.14 P_0^{1/2}\ell^{3/4}\kHz$.
Consider therefore the case of $P_0=1 \mW$ so that
$u_1=-253$.  For two-dimensional BECs a characteristic value for
the many-body parameter is $\eta\approx 10^{-3}$
\cite{PetHolShl00,BayTan98}.
Then for $N=10^5$ and $\ell=6$ we have
$\Delta r_6=1.9 \um$, and $\Omega_6=2\pi\times 0.54\kHz$.
This is not a particularly tight ring but tighter confinement
can be achieved by increasing the azimuthal mode index.
Figure 3 shows the variation of the oscillator frequency
$\Omega_\ell/2\pi$ in kHz (solid line), and $\Delta r_\ell$
in microns (dashed line) both versus $\ell$ for $u_1=-253, \eta N=100$,
and displays the advantage of using high azimuthal mode index.
We note that it is difficult to produce LG modes with very high azimuthal
index ($\ell > 6$) due to decreasing mode purity from holographic
generation.  However, other techniques exist for generating narrow annular
beams, e.g. using an axicon  \cite{ManekOG98}.
Typically the radial spread of such beams is very small, thus they
would allow one to explore the region of tight radial confinement seen
for LG beams with $\ell \ge 20$ (see Fig. 3).
\section{Loading a toroidal optical dipole trap}
Having established that LG beams can provide toroidal traps for
2D BECs the issue arises of how to load atoms into the ground state
of the trap.  It does
not seem feasible to condense directly in the toroidal trap as it does
not lend itself to evaporative cooling and the laser field could lead
to heating.  The solution is then to load into the toroidal trap from
an existing BEC, e.g. a magnetic trap or sheet dipole potential trap
\cite{Dur99}.  Here we consider loading of a toroidal trap from an
initial 2D ring BEC, for example using a magnetic trap
with a blue-detuned laser piercing its center \cite{DavMewAnd95},
and also from a
centrally peaked BEC as is the case for conventional harmonic traps.
The calculations presented here demonstrate that it is in principle
possible to load toroidal traps and we use illustrative examples which
highlight the issues involved.
\subsection{Types of initial condition}
We have numerically solved the GPE (\ref{GPeq4})
for a variety of initial conditions with the exact optical dipole
potential in Eq. (\ref{Uell}).
We consider the situation that for
$t<0$ the BEC is in the ground state of a prescribed
potential which is turned off for $t>0$ and the toroidal LG trap is
turned on.  To model this situation we have considered a variety of
initial conditions at $t=0$ and their subsequent evolution.  Rather
than dwelling on a specific potential model for $t<0$
we consider the super-Gaussian initial macroscopic wave functions
\begin{equation}
\psi(r,0) = {\cal N} e^{-(r-r_{\text{peak}})^m/w^m}  ,
\end{equation}
where ${\cal N}$ is a normalization constant, $w$ is the width of
the initial wave function, $m\ge 2$ is the order of the super-Gaussian,
$m=2$ being the usual Gaussian, and $r_{\text{peak}}$ is the
displacement of the density peak away from the origin: for
$r_{\text{peak}}=0$ we have a centrally peaked initial density,
$r_{\text{peak}}=r_T$ gives an initial ring BEC with
its peak at the toroid radius.  As $m$ increases
the initial condition becomes more top-hat like, representative of
the broadening due to repulsive many-body effects.  For the numerics
presented here we set $m=8$.

We have numerically solved the GPE (\ref{GPeq4})
for a variety of initial conditions with the exact optical dipole
potential in Eq. (\ref{Uell}) subject to the above initial conditions
using the split-step beam propagation method \cite{FleMorFei76}.
In implementing this scheme we have
included absorbing boundary conditions at the edge of the numerical grid,
thereby simulating losses due to radially outward going atoms.
We tested that the numerical results presented were not sensitive to the
placement of the absorber.  Thus, in our simulations the norm $n(t)$ in
Eq. (\ref{Norm}) of the macroscopic wave function is not conserved
at unity.  However, the norm $n(t)\le 1$ is actually a measure of the
fraction of initial atoms that are captured in the toroidal trap.

For most of the simulations presented we chose $\ell=6$ as this
is characteristic of the LG beams that can be generated reliably
experimentally at present \cite{CliffordACD98}.
\subsection{Initial ring BEC}
First we will discuss the loading of the toroidal optical dipole trap
with a BEC that is already ring shaped.
Although the BEC has already the desired shape, this transfer from a
magnetic-based ring trap to a toroidal trap is still of importance as
this trap can confine multi-component condensates in different Zeeman
sublevels. For an initial ring BEC the ground state of the toroidal trap
can be excited by reasonably matching the initial and ground
state wave functions.
Figure~\ref{RingBEC:fig} shows two examples of the computed dynamics for an
initial ring BEC formed in Na with $u_1=-25.3$ ($P_0=0.1 \mW$),
and $\eta N=50$.
For these parameters and $\ell=6$ the Thomas-Fermi theory of the last
section gives a ring width of $\Delta r_6=3.2\um$.
In Fig.~\ref{RingBEC:fig}(a), which shows a gray-scale plot of the 2D
atomic density $|\varphi(x,0,t)|^2$ versus time in ms
along the horizontal axis and radius $x$ in microns along the vertical axis,
we used $\ell=6$, $w=4\um$, and after a transient the
density profile settles down to a steady ring of smaller width
than the input.  The transient
involves expansion of the atoms outward from the ring and also
inwards towards the origin, and the peak at the origin
seen for $t\approx 3.2$ ms is due to interference between the inwardly
propagating circular atomic waves.  The outward propagating atomic waves
leads to an effective loss mechanism which allows for the damping
of the initial wave function towards the ground state toroidal BEC.
The final ring contains $n\approx 90\%$ of the initial atoms.
Figure \ref{RingBEC:fig}(b) shows a gray-scale
plot of the 2D atomic density for the same
parameters as \ref{RingBEC:fig}(a)
except $\ell=2$, for which $\Delta r_2=5.6\um$, so that the
input with $w=2.5\um$ is now half the expected ring width.
After a considerably larger transient than in
Fig. \ref{RingBEC:fig}(a) the density
again settles down but now to a larger ring than the input
containing about $n\approx 80\%$ of the initial atoms.
Notice, however, that for $t>15$ ms the atomic density
shows a further transient, and this is caused by weak atomic
waves caught around the origin that start to leak out, as
revealed by detailed examination of the data.

%
For these simulations we have used a rather
low laser power $P_0=0.1 \mW$, so that the toroidal trap depth is
only about $13 \nK$ for $\ell=6$.
We have chosen such a loosely bound toroidal trap as an illustrative
example of loading since tighter traps (larger $|u_1|$) are generally
easier to excite by suitably matched initial ring BECs as the ground
state and next excited state have a larger energy separation.
Thus the present example shows that there is considerable robustness
in loading from initial ring BECs, as expected intuitively.
One might expect such shallow traps to be very susceptible to heating
and losses induced by noise and background gas collisions.
However, the heating due to position and intensity noise increase with
the fourth and second power of the trap frequency $\Omega_\ell$,
respectively, and heating due to background gas collisions is
proportional to the square of the trap depth \cite{SavHarTho97,BaliOGGT99}.
Therefore our shallow traps (with low trap oscillation frequencies and
small trap depth) should be relatively immune to these detrimental effects.
%

\subsection{Initial centrally peaked BEC}
Figure~\ref{PeakedBEC:fig} shows representative 2D atomic density
plots from simulations of loading from an initial centrally
peaked ($r_{\text{peak}}=0$) Na BEC with
$m=8$,$u_1=-100$ ($P_0=0.40 \mW$), the corresponding trap depth being
$21.2 \sqrt{\ell}\nK$, and $\eta N=400$.  The plots
are for different initial BEC widths (a) $w=r_T=10\um$ with $\ell=6$,
(b) $w=14\um$ with $\ell=6$, and (c) $w=14\um$,
with $\ell=2$.  Case (a) shows very little signs of trapping in the
toroid which is centered at $r_T=10\um$, but (b) with an increased
BEC width of $w=14\um$ now shows substantial trapping with $N=90\%$ of
the initial atoms trapped.  What distinguishes these two case is that the
initial Hamiltonian in Eq. (\ref{Ham}), which we
evaluated numerically, is $H(0)/\hbar\omega_T=148$ in case (a) and
$H(0)/\hbar\omega_T=-16.6$ for case (b).  The Hamiltonian has three
contributions, the kinetic energy which is positive, the optical
dipole potential which for a red-detuned laser is negative, and
the nonlinear term due to many-body repulsion which is positive.
As we simulate the effects of atom losses due to outward propagating
atoms using the absorbing boundary, the atom number
decreases and so typically does the Hamiltonian
as the initial BEC evolves.  Physically, as we consider
red-detuned traps that have
negative optical dipole potentials which go to zero away from the trap,
atoms with energies above the trap energy of zero will tend to be lost
from the toroidal trap.
For case (a) the initial Hamiltonian or
energy is above the trap energy and the atoms fly above the
toroidal trap under the influence of the repulsive many-body effects.
Indeed the initial density profile with $w=10\um$ overlaps the
toroidal trap very little, leaving only positive contributions to
the Hamiltonian in Eq. (\ref{Ham}).
In contrast, for the wider initial BEC in
Fig.~\ref{PeakedBEC:fig}(b) the initial
density profile overlaps the toroidal trap giving a net Hamiltonian
less than the trap energy, so
trapping becomes possible.  We remark that the requirement that the
initial Hamiltonian be less than the trap energy
is a necessary but not sufficient
condition for trapping, as there is still dependence on the initial
density profile.  For example, Fig.~\ref{PeakedBEC:fig}(c)
is the same as in ~\ref{PeakedBEC:fig}(b)
except $u_1=-253$ ($P_0=1\mW$), the corresponding trap depth being
$130 \nK$ for $\ell=6$.  In this case $H(0)/\hbar\omega_T=-216$,
well below the trap energy,
but Fig.~\ref{PeakedBEC:fig}(c) shows that the atoms
do not cleanly load into the ground state of the toroidal trap,
but rather the density displays undamped oscillations on the time
scale of the simulation.  This oscillatory behavior is typically
a feature for deeper traps.  We have begun to explore methods of
smoothing the loading of deeper toroidal traps, such as tapering
the turn-on of the optical potential and phase-imprinting the
initial BEC so that it moves towards a ring. These methods do
smooth the loading process 
and we shall report on these issues along with the effects of beam
misalignments in a future publication.
\section{Summary and conclusions}
In this paper we  have shown that Laguerre-Gaussian beams provide
a flexible means for forming toroidal optical dipole traps in 2D
atomic BECs, and that the toroidal traps can be loaded from
initial conditions representative of conventional magnetic traps.
It remains to be seen how tight the atoms can be confined in these
rings, and detailed numerical studies are underway. This work is
 a first step towards developing toroidal traps which shall act
as mesoscopic rings for atoms for a variety of basic and applied studies
\cite{Rok97,JavPaiYoo98,BusAng99,PetYou99,BenRagSme99,GelGol00,IsoNakOhm00}.
For example, once a BEC is prepared in its ground state
of the toroidal trap a vortex state of variable angular momentum
can be excited using Raman coupling (involving a second LG beam)
to another Zeeman sublevel \cite{MarZhaWri97}, hence allowing for
studies of ring vortices and persistent currents on a torus
\cite{Rok97,JavPaiYoo98,GelGol00}. Also,
as our toroidal trap does not involve a magnetic trap it allows
for studies of multi-component BECs trapped on a ring \cite{IsoNakOhm00}.
In the limit of small number of atoms there is also the possibility
of realizing a Tonk's gas of impenetrable bosons on a ring
\cite{Ton36,Ols98,PetShlWal00}, which
has been predicted to exhibit dark solitons \cite{GirWri00}.
Furthermore, by looking at higher-order LG modes with different
radial index $p$, and hence multiple concentric rings, we can
create coaxial toroidal traps for the BECs, hence allowing for
radial tunneling between condensates.  In particular, multiple
rings could create a circular grating for atoms which could in
principle act as the feedback mechanism for a 2D atom laser, by
analogy to circular grating optical lasers \cite{ErdHal90}.
On the applied side,
once a ring BEC is formed we can further pierce it with
blue-detuned lasers at positions along its perimeter to form
tunnel-junctions akin to superconducting links.  One can then
envision a range of sensors, e.g. rotation, based on the
sensitivity of the tunneling current to any perturbation of the
system.  In principle, large rings could be made by first loading
a small ring and then adiabatically expanding the LG light mode
incident on the BEC, hence very sensitive matter-wave sensors of
inertial forces could result. \vspace{0.5cm}

\noindent EMW was supported in part by the Office of Naval
Research Contract No. N00014-99-1-0806, and the Department of Army
Grant No. DAAD 19-00-1-0169. KD acknowledges the support of the UK
Engineering and Physical Sciences Research Council.

\begin{figure}
\caption{Our basic model comprises a two-dimensional BEC which is
illuminated by a red-detuned LG beam traveling along the $z$-axis
and which acts as a toroidal trap.}
\end{figure}

\begin{figure}
\caption{Normalized optical dipole potential $U_\ell(r)/|U_1|$ versus
$r/r_T$ for various azimuthal mode indices $\ell$ for fixed laser power
and toroid radius.  For the red-detuning assumed here, as $\ell$ increases
the toroidal trap becomes deeper and tighter.}
\end{figure}

\begin{figure}
\caption{Variation of the oscillator frequency
$\Omega_\ell/2\pi$ in kHz (solid line), and width $\Delta r_\ell$
in microns (dashed line) both
versus $\ell$ for $u_1=-253$, $\eta N=100$.}
\end{figure}

\begin{figure}
\caption{Loading of a toroidal trap from an initial ring BEC with
$r_{\text{peak}}=10\um$ and $m=8$: Gray scale plots of the evolution of
the atomic density $|\varphi|^2$
for $u_1=-25.3$, $\eta N=50$, and (a) $w=4\um$ with $\ell=6$,
(b) $w=4$ with $\ell=2$.}
\label{RingBEC:fig}
\end{figure}

\begin{figure}
\caption{Loading of a toroidal trap from an initial super-Gaussian
with $m=8$: Gray scale plots of the evolution of the atomic density
$|\varphi|^2(x,0,t)$ for $u_1=-100$, $\eta N=400$, $\ell=6$, and
(a) $w=r_T=10\um$ with $H(0)/\hbar\omega_T=148$,
(b) $w=14\um$ with $H(0)/\hbar\omega_T=-16.6$.
(c) is for the same parameters as (b)
except $u_1=-253$ giving $H(0)/\hbar\omega_T=-216$.}
\label{PeakedBEC:fig}
\end{figure}
\newpage
\includegraphics*[width=0.6\columnwidth]{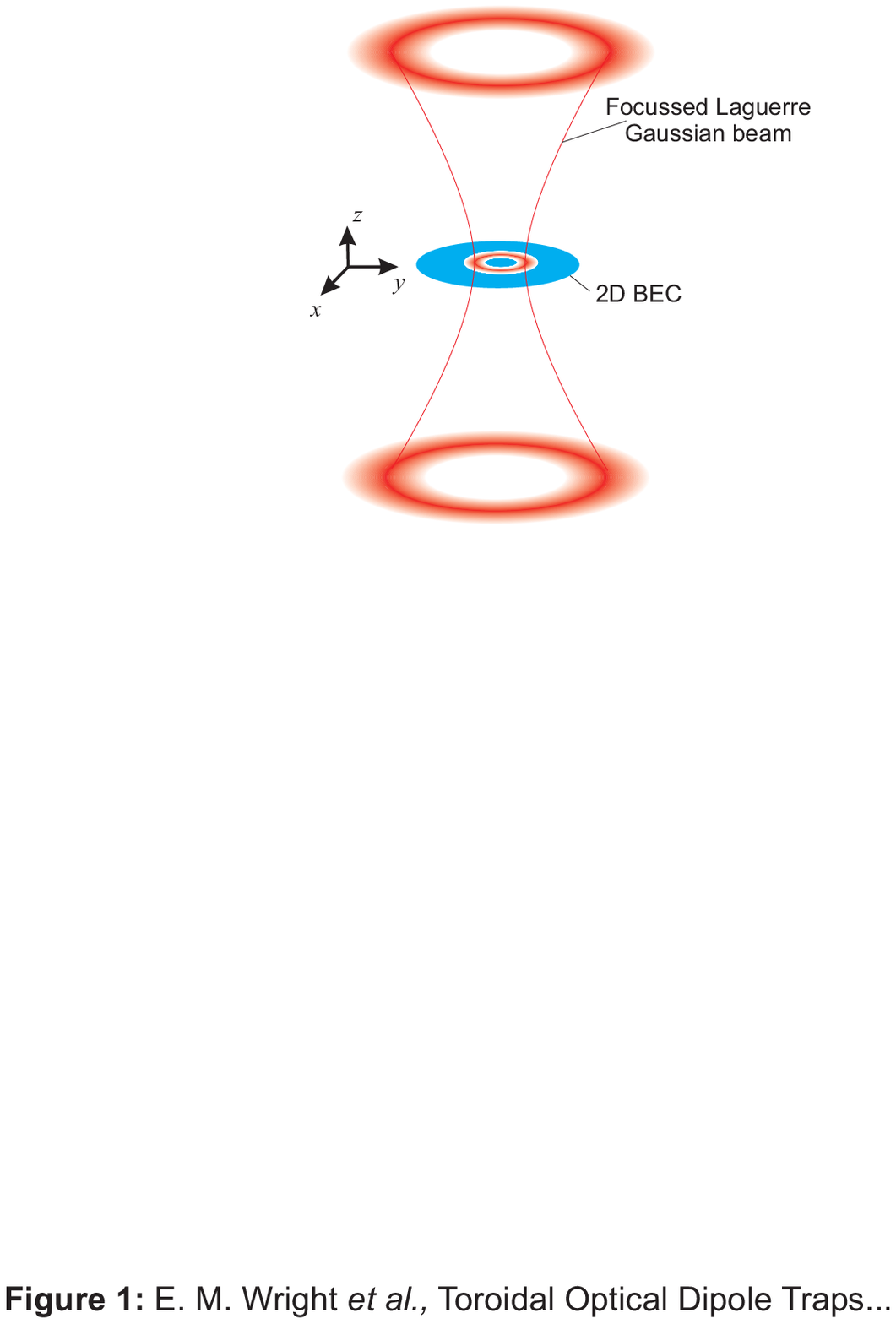}
\newpage
\includegraphics*[width=1.0\columnwidth]{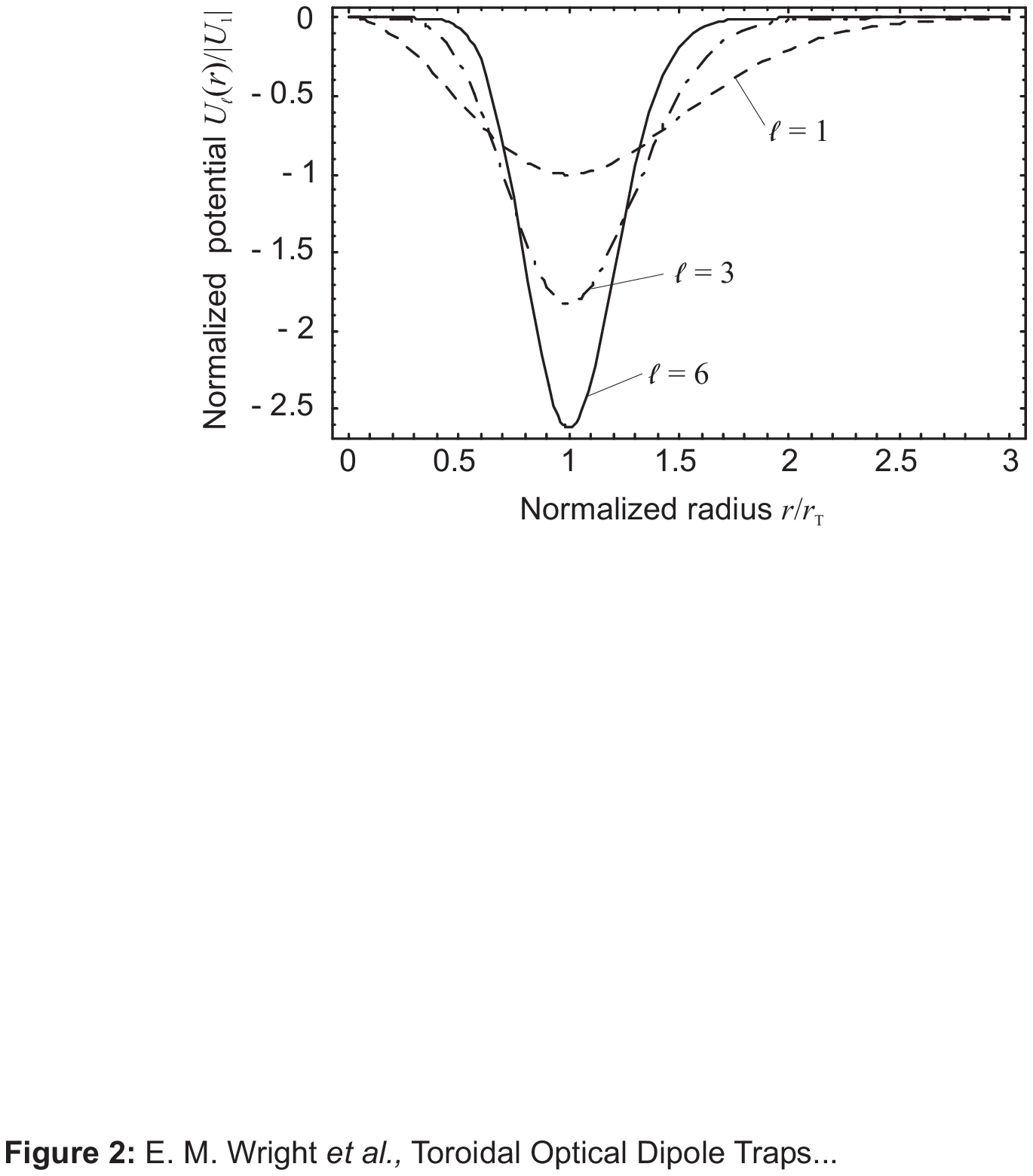}
\newpage
\includegraphics*[width=1.0\columnwidth]{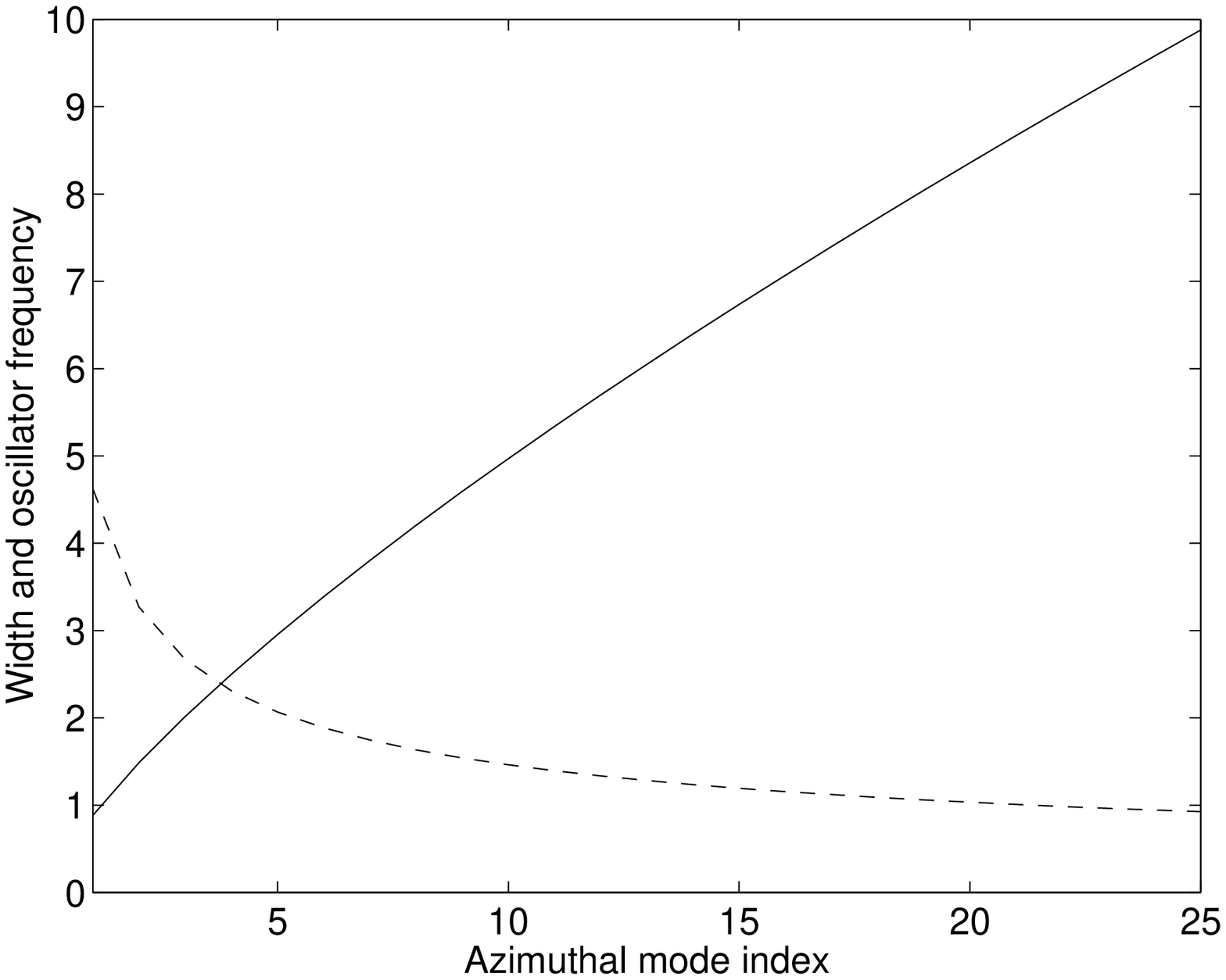}
\begin{center}
{\bf Figure 3}
\end{center}
\newpage
\includegraphics*[width=1.0\columnwidth]{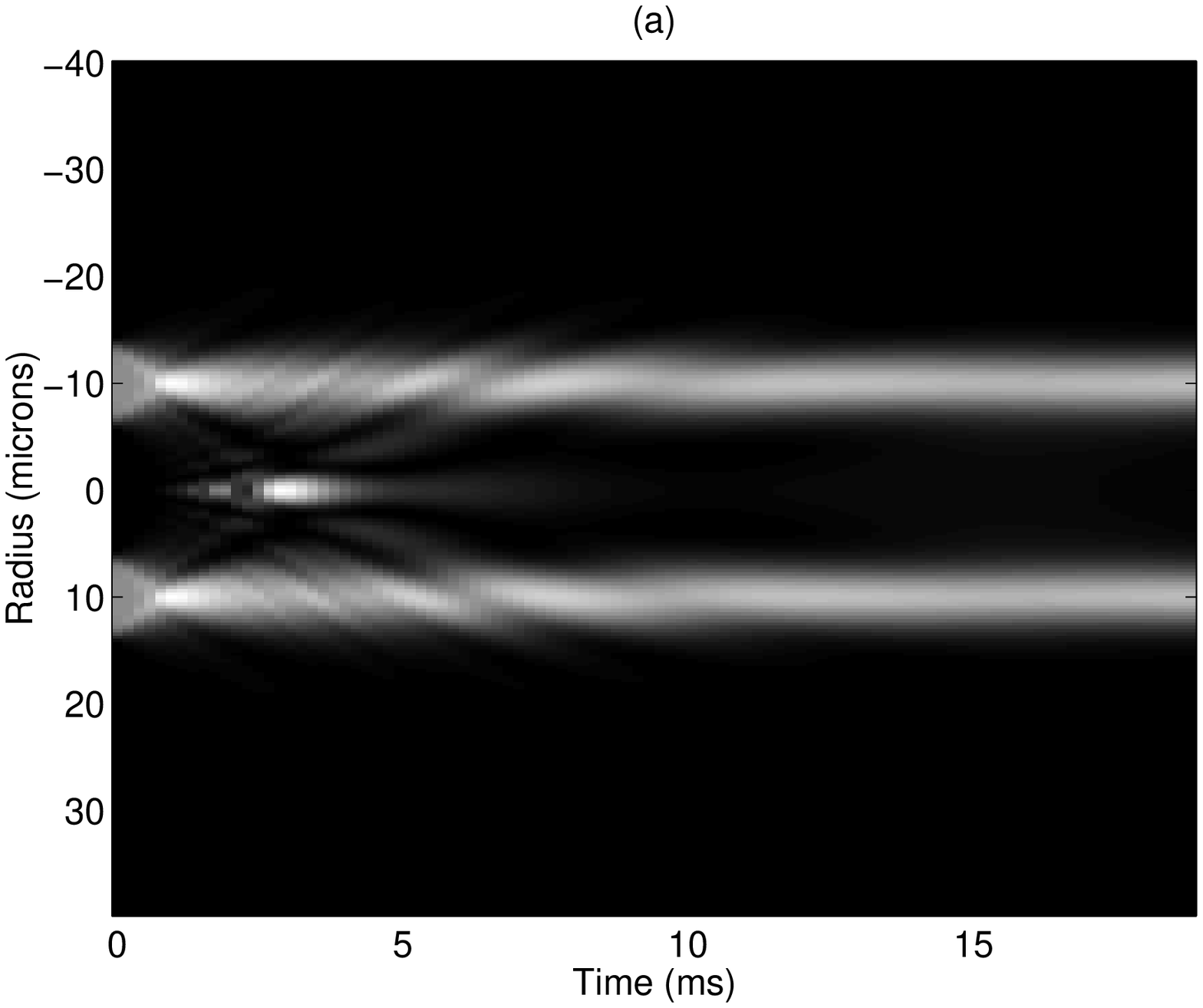}
\begin{center}
{\bf Figure 4(a)}
\end{center}
\newpage
\includegraphics*[width=1.0\columnwidth]{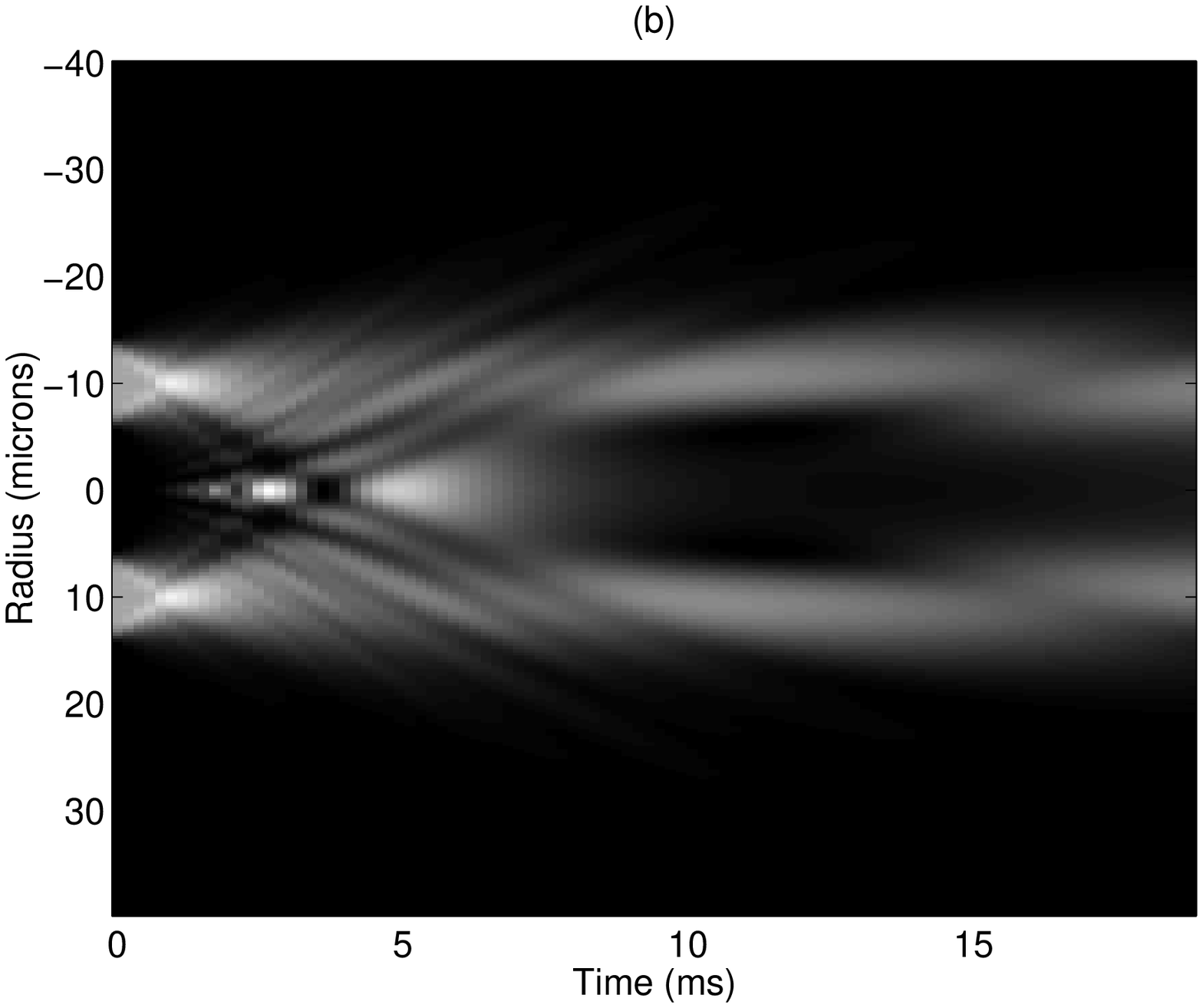}
\begin{center}
{\bf Figure 4(b)}
\end{center}
\newpage
\includegraphics*[width=1.0\columnwidth]{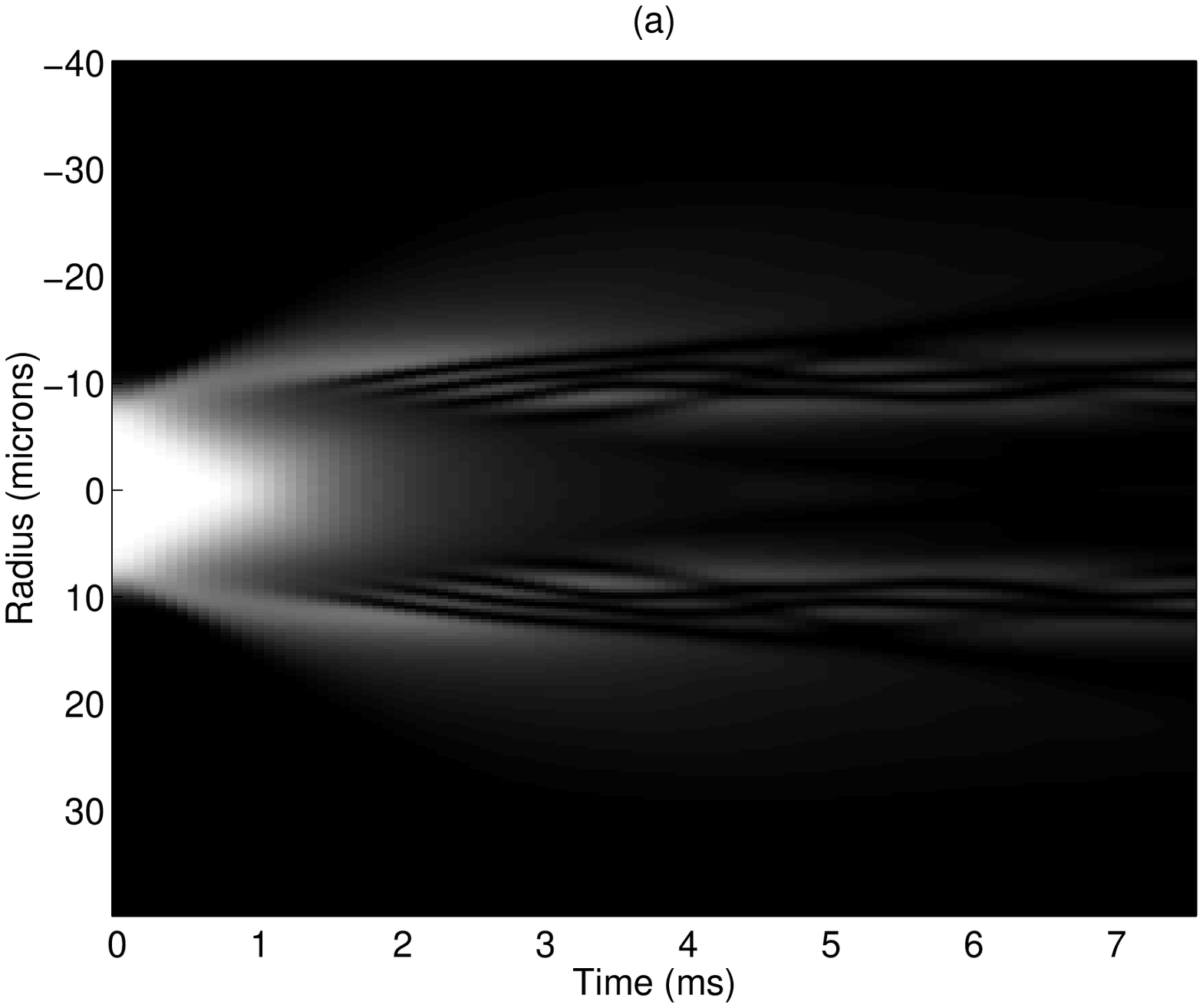}
\begin{center}
{\bf Figure 5(a)}
\end{center}
\newpage
\includegraphics*[width=1.0\columnwidth]{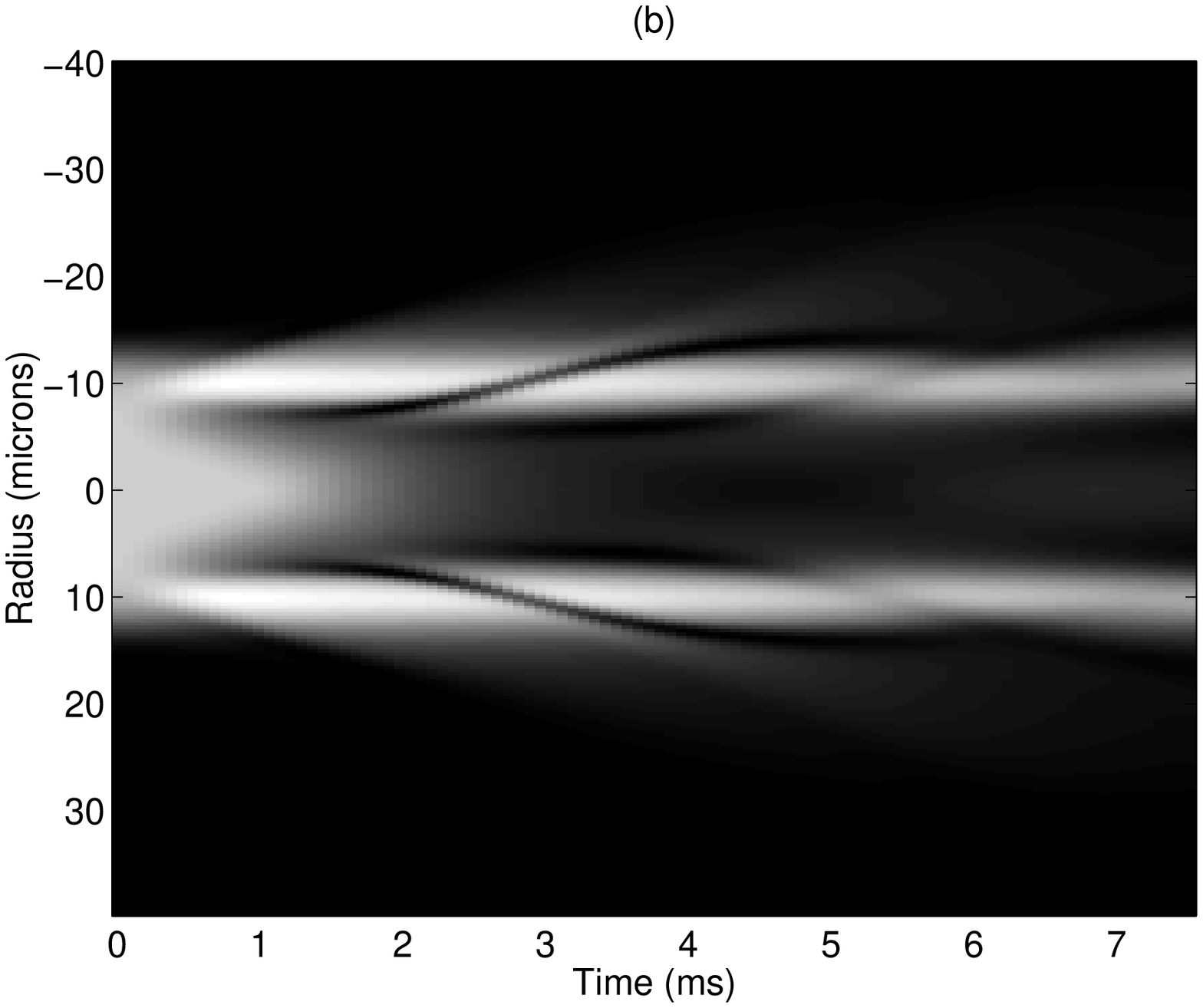}
\begin{center}
{\bf Figure 5(b)}
\end{center}
\newpage
\includegraphics*[width=1.0\columnwidth]{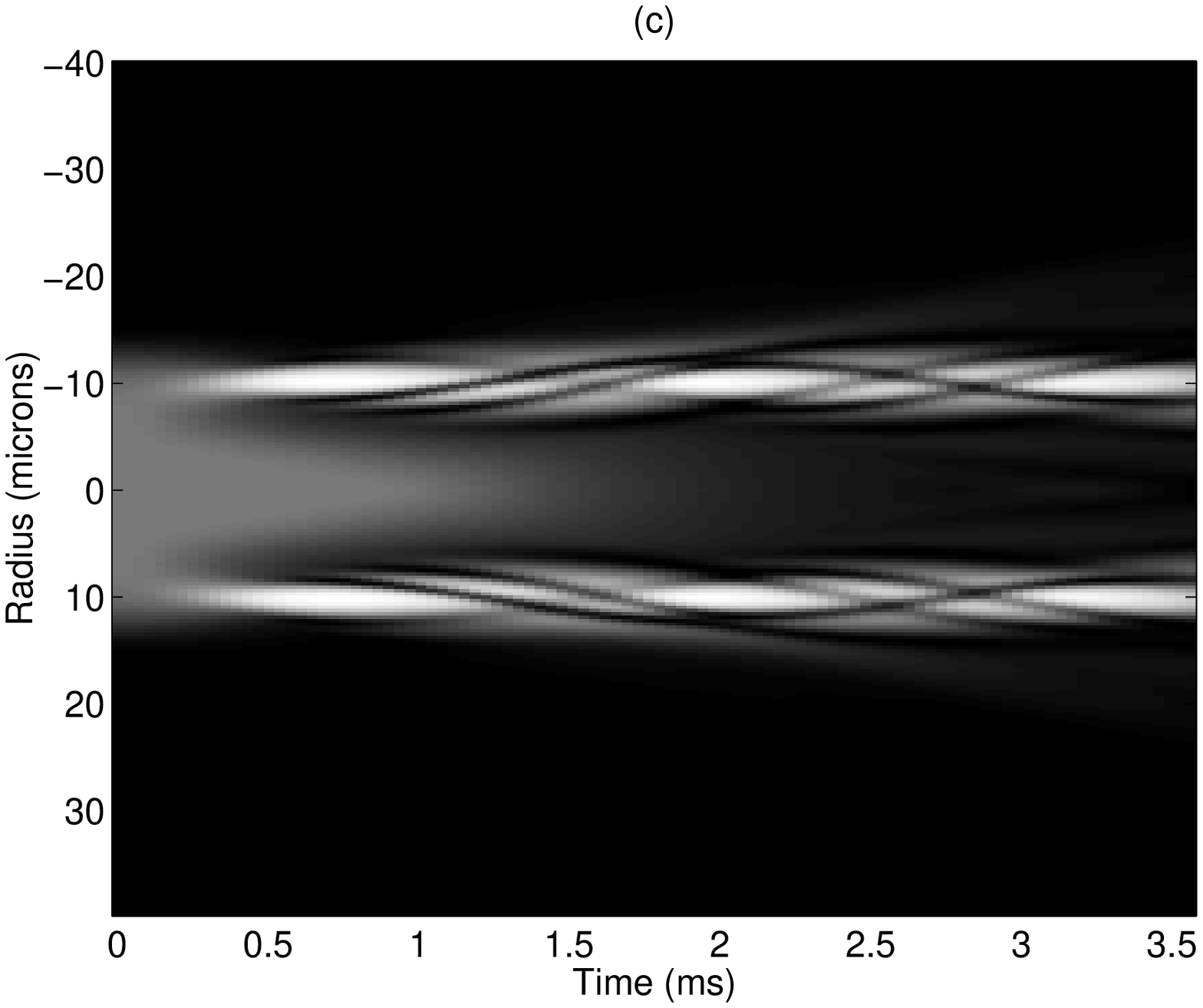}
\begin{center}
{\bf Figure 5(c)}
\end{center}
\end{document}